\documentclass[sigconf]{acmart}
\usepackage{color}
\usepackage{graphicx}
\usepackage{multirow}
\usepackage{subfigure}
\usepackage{hyperref}
\usepackage{amsmath}

\AtBeginDocument{%
  \providecommand\BibTeX{{%
    \normalfont B\kern-0.5em{\scshape i\kern-0.25em b}\kern-0.8em\TeX}}}

\setcopyright{acmcopyright}
\copyrightyear{2021}
\acmYear{2021}
\acmDOI{0/0}

\acmConference[DeepSpatial '21]{DeepSpatial '21: 2nd ACM SIGKDD Workshop on Deep Learning for Spatiotemporal Data, Applications, and Systems}{August, 2021}{Virtual Conference}
\acmBooktitle{DeepSpatial '21: 2nd ACM SIGKDD Workshop on Deep Learning for Spatiotemporal Data, Applications, and Systems, August, 2021, Virtual Conference}
\acmPrice{15.00}
\acmISBN{978-1-4503-XXXX-X/18/06}

\begin{document}

\title{Spatio-temporal Parking Behaviour Forecasting and Analysis Before and During COVID-19}

\author{Shuhui Gong}
\affiliation{%
 \institution{School of Information Engineering,}
 \institution{China University of Geosciences, Beijing}
 \city{Beijing}
 \country{China}
}
\email{shuhui.gong@cugb.edu.cn}

\author{Xiaopeng Mo}
\affiliation{%
 \institution{School of Computer Science,}
 \institution{University of Nottingham Ningbo China}
 \city{Ningbo}
 \country{China}}
\email{slyxm1@nottingham.edu.cn}

\author{Rui Cao}
\orcid{0000-0002-1440-4175}
\authornote{Corresponding author.}
\affiliation{%
 \institution{Dept. of LSGI \& SCRI,}
 \institution{The Hong Kong Polytechnic University}
 \city{Hong Kong}
 \country{China}
}
\email{rcao@outlook.com}

\author{Yu Liu}
\affiliation{%
\institution{Institute of Remote Sensing and Geographical Information Systems,}
\institution{Peking University}
\city{Beijing}
\country{China}}
\email{liuyu@urban.pku.edu.cn}

\author{Wei Tu}
\affiliation{%
 \institution{GD Key Lab. of Urban Informatics \& Dept. of Urban Informatics,}
 \institution{Shenzhen University}
 \city{Shenzhen}
 \country{China}}
 \email{tuwei@szu.edu.cn}

\author{Ruibin Bai}
\affiliation{%
 \institution{School of Computer Science,}
 \institution{University of Nottingham Ningbo China}
 \city{Ningbo}
 \country{China}}
\email{ruibin.bai@nottingham.edu.cn}


\begin{abstract}
  Parking demand forecasting and behaviour analysis have received increasing attention in recent years because of their critical role in mitigating traffic congestion and understanding travel behaviours.
However, previous studies usually only consider temporal dependence but ignore the spatial correlations among parking lots for parking prediction. This is mainly due to the lack of direct physical connections or observable interactions between them. Thus, how to quantify the spatial correlation remains a significant challenge. To bridge the gap, in this study, we propose a spatial-aware parking prediction framework, which includes two steps, i.e. spatial connection graph construction and spatio-temporal forecasting.
A case study in Ningbo, China is conducted using parking data of over one million records before and during COVID-19.
The results show that the approach is superior on parking occupancy forecasting than baseline methods, especially for the cases with high temporal irregularity such as during COVID-19.
Our work has revealed the impact of the pandemic on parking behaviour and also accentuated the importance of modelling spatial dependence in parking behaviour forecasting, which can benefit future studies on epidemiology and human travel behaviours.
\end{abstract}




\maketitle

\section{Introduction}
With the rapid increase of private cars, finding a free parking space in cities becomes increasingly challenging. This lays a heavier burden on road traffic and further contributes to congestion \cite{yang2019deep}. 
Parking demand prediction can help adjust the parking services dynamically to improve the overall efficiency and has the potential to optimise the parking space finding process \cite{fiez2018data}.
Therefore, the problem is significantly important and has attracted growing attention.

Owing to the development of information and communication technologies (ICT), we have access to more and more digital parking data, which enable us to probe into the parking behaviour from historical time-series parking records.
Based on these available data, previous studies used various methods to predict parking occupancy, including support vector regressions (SVR) \cite{zheng2015parking}, clustering \cite{tamrazian2015my}, time series models \cite{liu2010unoccupied}, and neural networks \cite{alajali2017street,yang2019deep}.
Most of the methods only focused on leveraging temporal features of the historical records, without considering the spatial connections among parking lots. 
 
However, numerous studies have shown that passengers' travel behaviours have significant spatial correlations \cite{tu2018spatial,gong2019extracting}. For example, studies show that passengers would take up to 500-metre walking to their final destinations \cite{yue2012exploratory}. This suggests that residents may park their own cars in parking lots within tolerable distance of the destination, and walk to the destination after parking. The finding also implies the existence of spatial correlation between parking lots which locates near each other.

Furthermore, only considering temporal features will make the established model vulnerable to unexpected events that suffer from dramatic temporal pattern change, for example, the COVID-19, which first broke out in China and then spread globally in a short time \cite{world2020coronavirus}.
During the pandemic, compulsory restrictions were firmly implemented in China to control the respiratory infectious disease, including physical distancing and community containment measures, which significantly reduced public transport use and public gatherings \cite{zhou2020effects}.
These governmental policies dramatically changed the way people travel and parking.
Therefore, the emergencies would pose great challenge for parking behaviour forecasting, in which case historical temporal patterns would not be sufficient.

To address these issues, in this study, unlike previous studies only considering historical time series, we proposed a spatio-temporal parking prediction framework to predict parking occupancy in the next future, accounting for both temporal and spatial correlations. The framework includes two major steps: the first step is to construct a connection graph to model the spatial connections between parking lots, the second step is spatio-temporal parking forecasting using historical time-series parking data and the generated connection graph.
We performed a case study in the city of Ningbo, China, using more than one million parking records of 136 parking lots from April 2019 to April 2020 (including parking data before and during the period of COVID-19).
The results demonstrate the superiority of our proposed framework in parking forecasting, especially during COVID-19 when the parking behaviours exhibited significantly temporal irregularity than normal days.

The rest of the paper is organised as follows.
Section \ref{sec:literature_review} reviews the related works.
Section \ref{sec:forecasting} presents the proposed spatial-aware parking forecasting framework.
Section \ref{sec:data} introduces the study area and data.
Section \ref{sec:analysis} presents the experimental results and analysis.
Finally, Section \ref{sec:conclusion} concludes the paper.

\section{Literature Review}
\label{sec:literature_review}
Here, we focused on reviewing the related works for parking occupancy prediction. In recent years, quite a few methods are  proposed to estimate parking occupancy. These methods can be divided into two categories: first is $statistical$ and $machine$ $learning$ methods, the other is $deep$ $neural$ $networks$ \cite{yang2019deep}.

The $statistical$ and $machine$ $learning$ approaches often use historical parking data to predict short-term parking occupancy \cite{pullola2007towards}. The methods include clustering \cite{tamrazian2015my}, Support Vector Regression (SVR) \cite{zheng2015parking}, Autoregressive Integrated Moving Average (ARIMA) \cite{burns1992econometric}, continuous-time Markov Chain model \cite{klappenecker2014finding}, etc. With the development of technology, models of deep neural networks are gaining popularity in the area of parking modeling and prediction. Ji et al. \cite{ji2014short} proposed a three layer wavelet neural network to predict short-term parking availability in off street garages; Vlahogianni et al. \cite{vlahogianni2016real} approached multi-layer perceptrons (MLPs) in predicting the parking occupancy; Ziat et al. \cite{ziat2016joint} adopted a representation learning method to simultaneously predict traffic state and parking occupancy. These studies largely improved the accuracy of the prediction results compared with the statistical method. They also consider the external factors that influenced the drivers' parking behaviours, such as parking duration, traffic congestion, weather, etc. However, most previous studies did not consider the location correlations among the parking lots, which would lack knowledge to analyse travel behaviours by private cars, and would make the prediction accuracy lower as well.

Graph convolutional neural network (GCNN) is a machine learning technology that uses the graph spectral theory to filter the signals on localised sub-graphs.
The filtered signals are considered as the features for the neural networks. In recent years, GCNN has been used on spatial correlations modelling and prediction. Zhao et al. \cite{zhao2019t} integrated GCN and GRU for short-term traffic forecasting. Yang et al. combined GCN with RNN to predict real-time on-street parking \cite{yang2019deep}. Although this work consider both spatial and temporal correlations, it only consider on-street parking lots, which is directly influenced by the traffic flow. These occupancies could be much easier to predict than general parking lots. Moreover, it use real time traffic flow instead of historical data to make prediction.

Based on the background, in this study, we proposed a new spatial-temporal parking model that can capture the temporal and spatial features from the parking occupancy data, and can be used for parking services prediction based on urban design.

\section{Spatio-temporal Parking Behaviour Forecasting}
\label{sec:forecasting}

\subsection{Problem Formulation}
The problem of parking behaviour (measured by parking occupancy in our case) forecasting can be regarded as learning of the mapping function $f$ using the locations of parking lots and associated time series observations $\{X_i\}_{i=t-n}^t$, and then predicting the parking occupancy in the next $T$ time periods, as formulated as follows:
\begin{equation}
\left[X_{t+1}, \cdots, X_{t+T}\right]=f\left(L, \left[X_{t-n}, \cdots, X_{t-1}, X_{t}\right]\right)
\end{equation}
where $n$ and $T$ are the lengths of the historical and forecasting time series, respectively.
$L=\{(\lambda_i, \phi_i)\}_{i=1}^{N}$, in which $\lambda$ and $\phi$ represent longitude and latitude, respectively.

Most previous studies \cite{burns1992econometric,pullola2007towards} ignore the location information $L$ of parking lots and their associated spatial correlation due to the lack of direct physical connections or observable interactions between them.
However, parking lots are not absolutely independent, the state of one parking lot may well impact the others, and usually the nearer ones are more likely to be influenced.
The challenge is how to model the spatial correlation among them.
To address this problem, we propose a spatial-aware parking prediction framework to leverage both temporal and spatial dependence.

\subsection{Methodology}
\begin{figure*}
\centering
\includegraphics[width=0.98\textwidth]{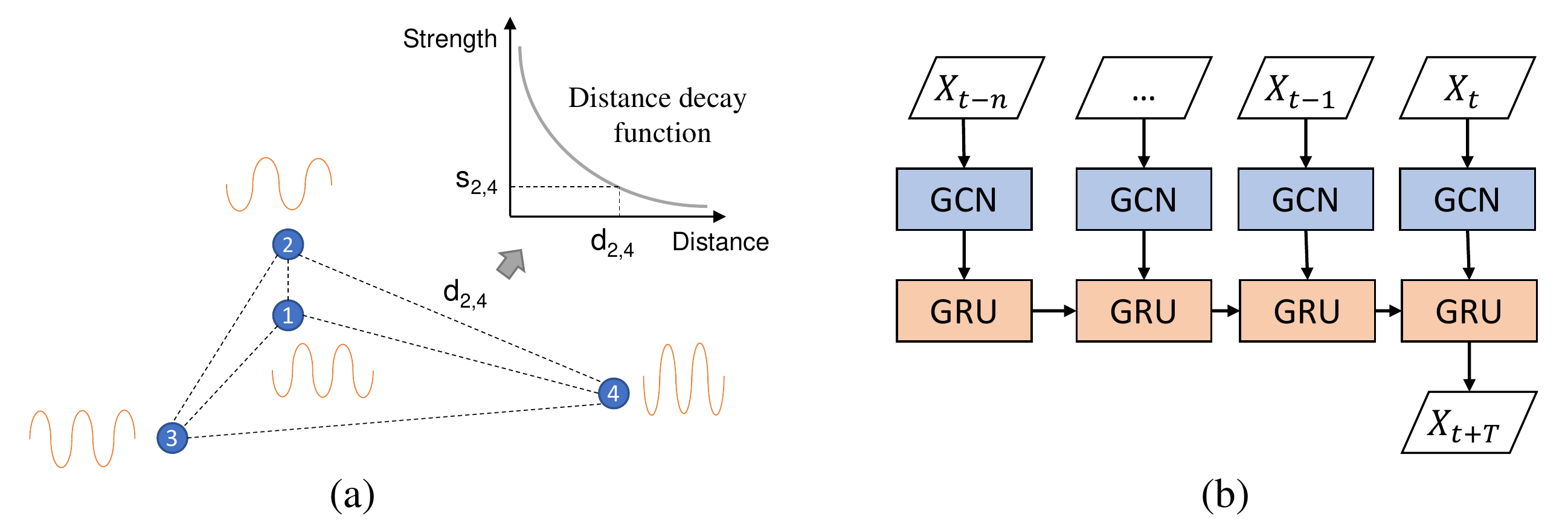}
\vspace{-0.5cm}
\caption{Illustration of the proposed framework, which includes two steps: (a) connection graph construction and (b) spatio-temporal forecasting.}
\label{fig:method}
\end{figure*}

The proposed framework of spatio-temporal parking prediction is presented in Fig.~\ref{fig:method}.
First, we build the connection graph from different locations. Then, we use a temporal graph convolution network to make spatio-temporal forecasting.

\subsubsection{Connection Graph Construction:}
To model the spatial dependence of the parking lots, we propose to construct a weighted graph $G=(V, E)$ from locations $L$, as illustrated in Fig. \ref{fig:method}(a), where $V=\{v_i\}_{i=1}^{N}$ is the node set and $E=\{e_{i,j}\}_{i,j=1}^N$ is the edge set.
In the graph, each node $v_i$ denotes a parking lot and each edge $e_{i,j}$ represents the connection between node $v_i$ and $v_j$.
The weighted adjacency matrix $A$ can be used to indicate the topological links and the associated connection strength between the parking lots, where $A \in R^{N\times N}$, with the entry values $A_{i,j}\geq 0$, and higher value indicates stronger connection.


As we know, it is common that people select another nearby parking lot if their targeted one has no space for parking. Therefore, the closer parking lots usually present similar parking patterns and have stronger spatial correlation.
Based on this observation, unlike previous studies only considering connectivity \cite{zhao2019t}, we construct weighted adjacency matrix accounting for both connectivity and connection strength (measured by distance decay effect) to model the spatial dependence among parking lots.

Specifically, the weighted adjacency matrix can be formulated as follows:
\begin{equation}
A_{i,j} = \left\{\begin{array}{l}
d_{i,j}^{-\beta}, ~ \mathrm{if}~d_{i,j} \leq d_0\\ 
~~0, ~~~ otherwise
\end{array}\right.
\end{equation}
where $A_{i,j}$ reprsents the connection strength between the $i$-th and $j$-th parking lots, while $d_{i,j}$ is the distance between them.
$d_0$ is the distance threshold, which reflects people's tolerance of walking distance between the parking locations and their final destinations. 
$\beta$ is the distance decay factor. When $\beta=0$, the weighted adjacency matrix will reduce to normal binary adjacency matrix.


\subsubsection{Spatio-temporal Forecasting:}
We use T-GCN \cite{zhao2019t} as our spatio-temporal forecasting model for parking prediction, which mainly consists of two components: a graph convolutional network (GCN) \cite{kipf_semi-supervised_2017} and gated recurrent units (GRU). The structure of T-GCN is presented in Fig. \ref{fig:method}(b).

Firstly, for each time $t$, the input feature matrix $X_{t}$ and the weighted adjacency matrix $A$ are fed into a two-layer GCN $g$ to learn new features $X_t^s$, which have incorporated into spatial dependence from the locations of parking lots:
\begin{equation}
X_t^s = g(A, X_t)=\sigma(\hat{A} ReLU(\hat{A} X_t W_{0})W_{1})
\label{gcn}
\end{equation}
where $\hat{A}=\widetilde{D}^{-\frac{1}{2}}\widetilde{A}\widetilde{D}^{-\frac{1}{2}}$, $\widetilde{A}=A+I_{N}$, $\widetilde{D}=\sum_{j}\widetilde{A}_{ij}$;
$W_0$ and $W_1$ are the weight matrices of the first and second layer, while $\sigma$ and $ReLU$ are the activation functions.
%
Then, the new feature matrix series $\{X_i^s\}_{i=t-n}^{t}$ will be fed into GRU repeatedly to learn the temporal features. The updating process can be formulated as follows:
\begin{equation}
\begin{array}{l}
u_{t}=\sigma\left(W_{u}\left[X_t^s, h_{t-1}\right]+b_{u}\right) \\
r_{t}=\sigma\left(W_{r}\left[X_t^s, h_{t-1}\right]+b_{r}\right) \\
c_{t}=\tanh \left(W_{c}\left[X_t^s,\left(r_{t} * h_{t-1}\right)\right]+b_{c}\right) \\
h_{t}=u_{t} * h_{t-1}+\left(1-u_{t}\right) * c_{t}
\end{array}
\end{equation}
where $u_t$ and $r_t$ are the update and reset gates at time $t$, $c_t$ is the cell state and $h_t$ denotes the output at time $t$. $W$ and $b$ represent weights and biases.

In the training process, the goal is to minimise the error between the real parking occupancy and the predicted one. Back-propagation algorithm is used to update the parameters of the model.
After training, T-GCN can model both the spatial and temporal dependence and learn to predict future parking occupancy. Then, the trained model can be used for parking forecasting.

\section{Study Area and Data}
\label{sec:data}
\begin{figure}
\centering
\includegraphics[width=0.4\textwidth]{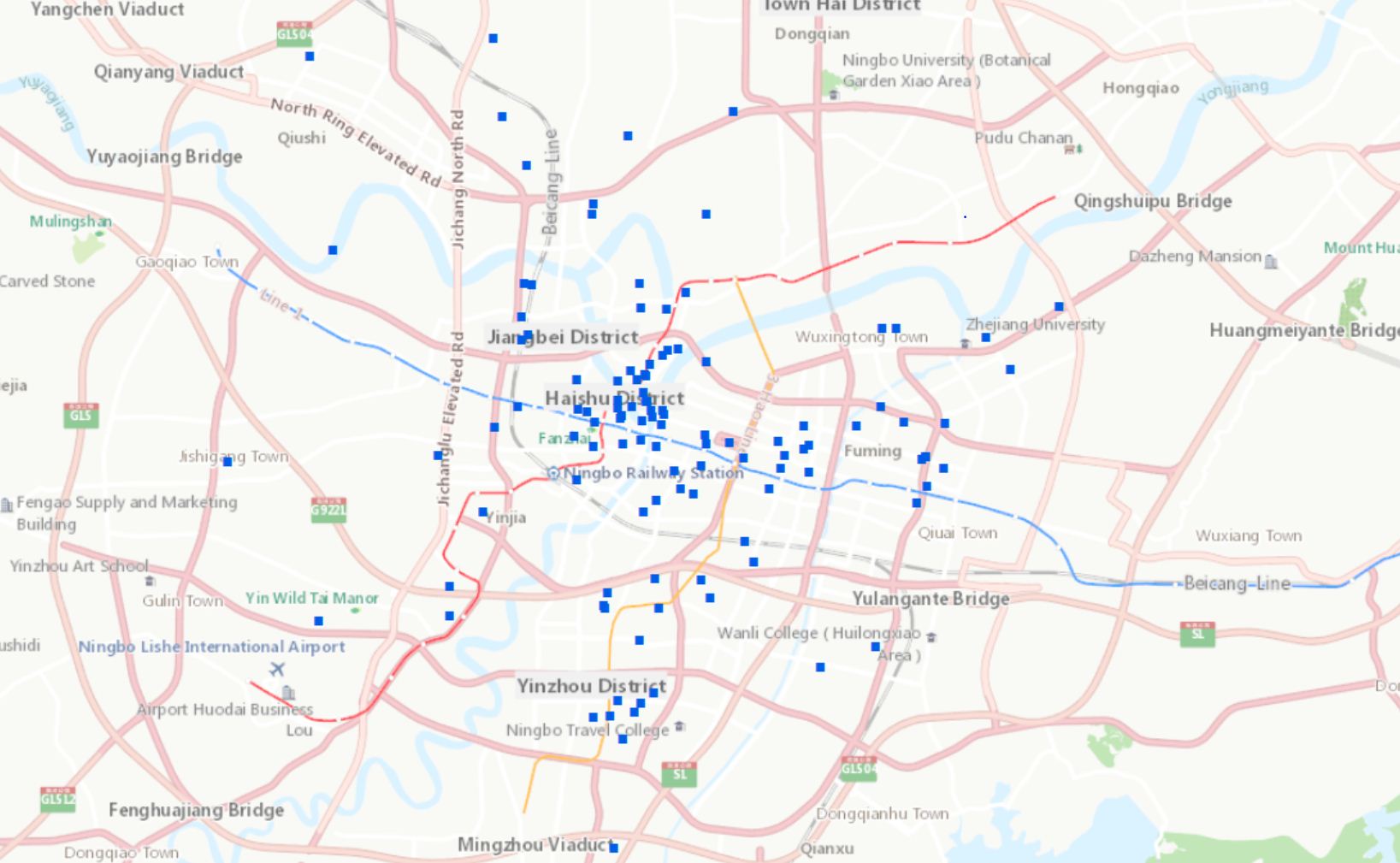}
\caption{Spatial distribution of the 136 parking lots (blue dots) in Ningbo, China.}
\label{fig:parking_location}
\end{figure}

We take Ningbo, China as our study area, which is a major port and industrial hub in the east of China and lies south of Shanghai, with a population of 7.6 million.
One-month parking data in 136 parking lots are used before and during COVID-19.
The spatial distribution of the parking lots are shown in Fig.~\ref{fig:parking_location}. As can be seen, the 136 parking lots occupy most of the areas in Ningbo.
Concretely, we collect one-week parking data in November 2019 (including 5 working days and 2 non-working days), and 17 days parking data during COVID-19 (including 11 working days and 6 non-working days from February 2020 to April 2020). The data include information of parking lots, such as name, location (longitude and latitude), open time, recording time, and parking occupancy.
To guarantee the data quality, we clean the data with two preprocessing steps: (i) for each parking lot, we only retain data once per hour by calculating average occupancy per hour; (ii) we only retain data of 12 hours, from 8 am to 8 pm before COVID-19, and 8 am to 6 pm during COVID-19 due to the data limitation.


\subsection{Spatio-temporal Parking Patterns}
\begin{figure*}
\centering
\includegraphics[width=\textwidth]{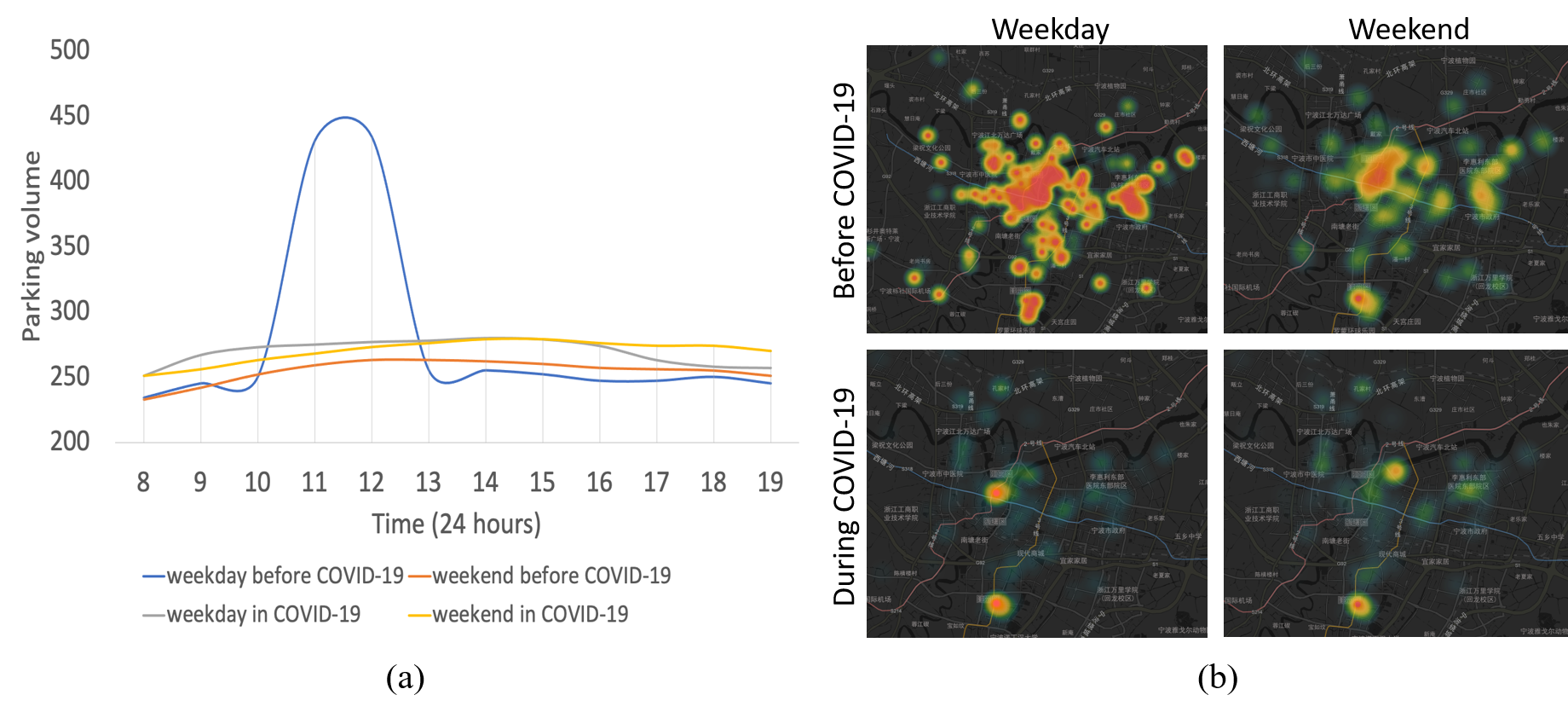}
\vspace{-0.8cm}
\caption{Average parking volume in (a) different time and (b) different locations.}
\label{fig:st-patterns}
\end{figure*}

We analyse the spatio-temporal parking patterns in Ningbo before and during COVID-19 by aggregating the parking volumes from the perspective of time and space, respectively. The results are shown in Fig. \ref{fig:st-patterns}.

As can be seen in Fig. \ref{fig:st-patterns}(a), We find that the parking volume in weekdays before COVID-19 has a distinct peak from around 11 am to 12 pm, which is quite unusual compared with other times. This suggests that most people prefer to park their private cars in the parking lots at noon, instead of driving them on the road. The finding is interesting but not surprising, since in most urban cities in China, there often tends to be less traffic at noon. 
In other times, the parking occupancy volume are relatively stable. The temporal patterns show that the parking behaviours have changed drastically during the COVID-19 period, which also implies that the normal commuting life has been deeply affected by the unexpected event.

Fig.~\ref{fig:st-patterns}(b) shows the heatmap of passengers' parking distribution at 8 am of weekday and weekend before and during COVID-19.
We can see that:
(i) Comparing the parking occupancy distribution before (upper row of Fig.~\ref{fig:st-patterns}(b)) and during COVID-19 (bottom row of Fig.~\ref{fig:st-patterns}(b)), we see that passengers' driving behaviours are much more active and diverse before COVID-19. More specifically, before COVID-19, parking lots in business buildings, schools, and entertainment areas have very high parking occupancy volume in weekdays. While on weekends, private cars are mostly located in city centres, around Tianyi Square for shopping, and the Old Bund for entertainment.
(ii) From the bottom row of Fig.~\ref{fig:st-patterns}(b), we see that passengers have very similar parking behaviours in different times during COVID-19. Most passengers' cars are parked in residential areas in weekday and weekend. In particular, Ronganfu residential community, and Tuyuan residential community in weekday (two hotspots in the bottom-left of Fig.~\ref{fig:st-patterns}(b)), and Ziyutai residential community, and Ronganfu residential community in weekend (two hotspots in the bottom-right of Fig.~\ref{fig:st-patterns}(b)).

\section{Experiments and Analysis}
\label{sec:analysis}
\subsection{Experiment Setup}
To verify our method, we split each dataset (before and during COVID-19) into two subsets. Specifically, we take first 80\% of the data for training, and the rest 20\% are used to evaluate the results.

To construct the weighted adjacency matrix, we use grid search to find the best decay factor $\beta$ within $[0, 2]$, which is empirically set to be 1.25.
The distance threshold $d_0$ is set to 500 metres (Euclidean distance) in our experiments since previous studies indicate that people would take up to 500-metre walking to their final destination \cite{yue2012exploratory}. We have also compared different thresholds to examine the impact of distance thresholds in the following experiments.

To train our framework, we empirically adjust and set the learning rate to 0.001, the batch size to 64, and the training epoch to 3000.
Besides, we also approached four other baseline methods (ARIMA, ANN, SVR, and GRU) for comparisons.
Three criteria (RMSE, MAE, MAPE) are used as evaluation metrics. For each criterion, larger value indicates greater error, and vice versa.

\begin{table*}
\centering
\caption{Overall parking occupancy prediction results before COVID-19.}
\label{tab:t-gcn_result2}
\resizebox{0.8\textwidth}{!}{
\begin{tabular}{c|ccccc|ccccc}
\hline
\multirow{2}{1.0cm}{Metric} & \multicolumn{5}{c}{Weekday}&\multicolumn{5}{c}{Weekend}\\
\cline{2-11}
&ARIMA & SVR&ANN&GRU&Ours&ARIMA & SVR&ANN&GRU&Ours\\
\hline
RMSE&0.12&0.11&0.10&0.07&\textbf{0.07}&0.08&0.08&0.07&0.06&\textbf{0.06}\\
MAE &0.11&0.10&0.08&0.05&\textbf{0.05}&0.07&0.07&0.07&0.07&\textbf{0.06}\\
MAPE & 21.50&24.52&30.55&11.12& \textbf{10.74}&18.52&19.23&23.46&12.77&\textbf{12.62}\\
\hline
\end{tabular}}
\end{table*}

\vspace{1cm}
\begin{table*}
\centering
\caption{Overall parking occupancy prediction results during COVID-19.}
\label{tab:t-gcn_result}
\resizebox{0.8\textwidth}{!}{
\begin{tabular}{c|ccccc|ccccc}
\hline
\multirow{2}{1.0cm}{Metric} & \multicolumn{5}{c}{Weekday}&\multicolumn{5}{c}{Weekend}\\
\cline{2-11}
&ARIMA & SVR&ANN&GRU&Ours&ARIMA & SVR&ANN&GRU&Ours\\
\hline
RMSE&0.10&0.25&0.10&0.13&\textbf{0.09}&0.89&0.62&0.60&0.29&\textbf{0.28}\\
MAE &0.08 &0.06&0.09&0.11&\textbf{0.06}& 0.89&0.73&0.59&0.14&\textbf{0.11}\\
MAPE & 36.66&47.90&29.14&13.10&\textbf{11.82}&47.90&25.89&36.00&13.38&\textbf{12.22}\\
\hline
\end{tabular}}
\end{table*}

 \begin{figure*}
\centering
\subfigure[Shipu restaurant]{\label{fig:shipu_result}
\includegraphics[width=0.8\textwidth]{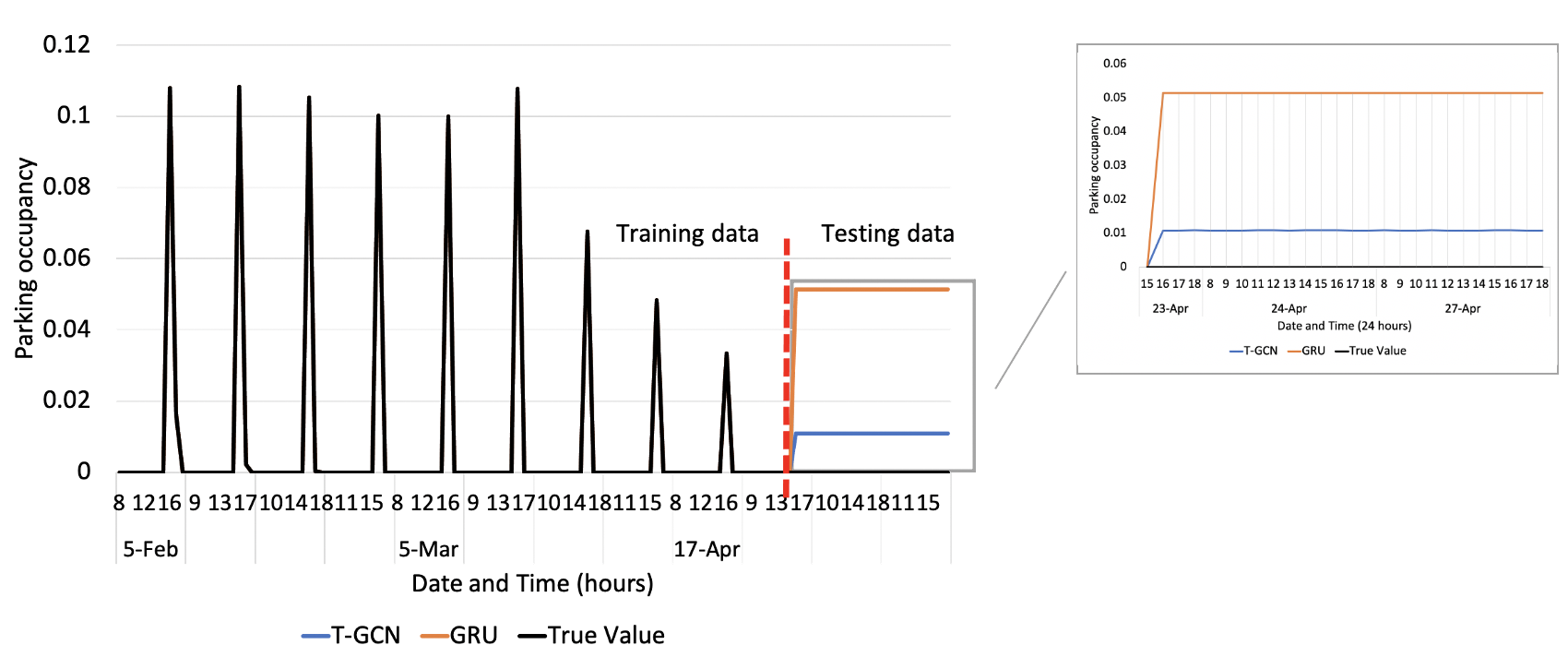}}
\subfigure[Shanjing shopping centre]{\label{fig:shanjing}
\includegraphics[width=0.8\textwidth]{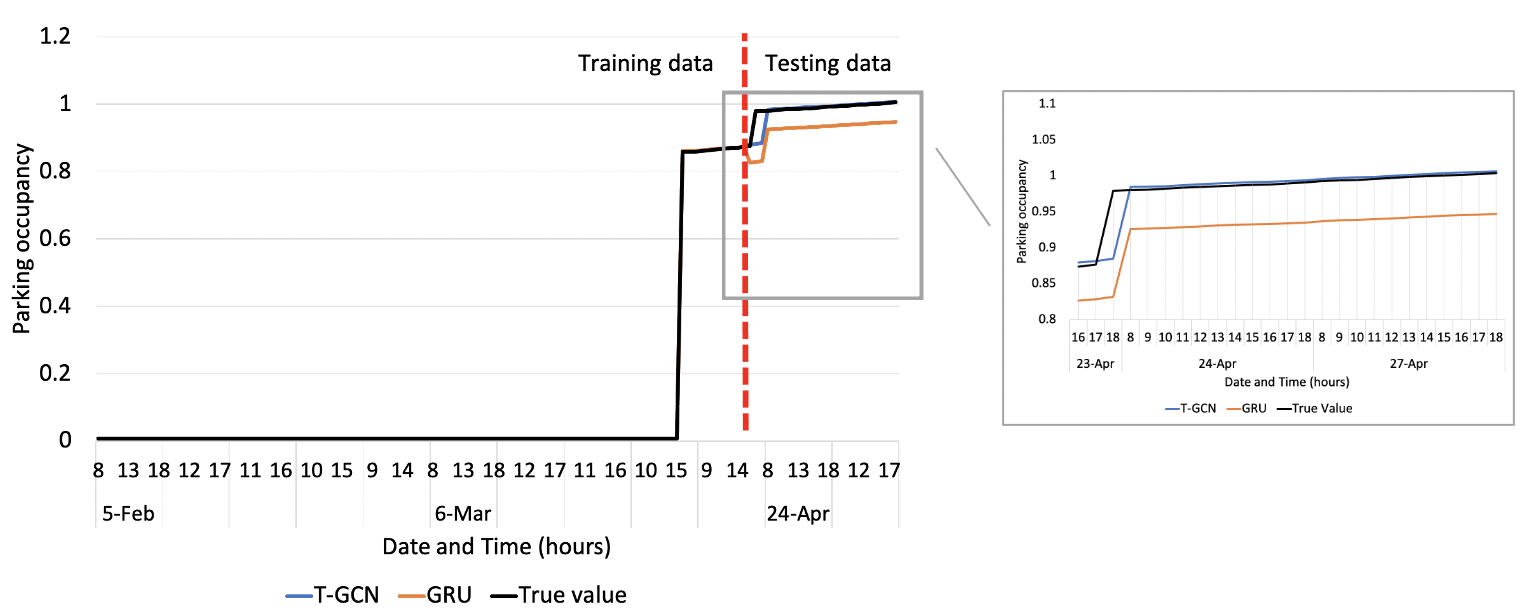}}
\vspace{-0.3cm}
\caption{Predicted parking occupancy of exemplar parking lots during COVID-19.}
\label{fig:parking_covid_time}
\end{figure*}

\subsection{Parking Forecasting Results}
\label{sec:result}
The overall parking occupancy prediction results using data from before and during COVID-19 are shown in Table~\ref{tab:t-gcn_result2} and Table~\ref{tab:t-gcn_result}, respectively.
We can see: (i) Our framework achieves the best results with the lowest errors compared with other methods in all the time periods (before and during COVID-19). In particular, our framework has higher prediction accuracy than GRU, which demonstrates the effectiveness of modelling spatial dependence among parking lots and further implies that parking behaviours in nearer parking lots have stronger spatial correlations since $\beta>1$.
(ii) In general, the parking prediction errors during COVID-19 is higher than that of normal days for both GRU and our framework (e.g. RMSE is 0.09 and 0.28 during the pandemic compared with 0.07 and 0.06 before COVID-19 for our framework). This implies that the parking behaviours are much more irregular during COVID-19 and it is harder for the models to make right predictions.
(iii) Our framework improves the prediction performance (over GRU) more significantly during COVID-19, with RMSE decrease of 0.04 and 0.01 during COVID-19 and limited improvement before COVID-19. This suggests that the modelling of spatial dependence contributes more with cases of higher time irregularity.

\begin{table*}[h!]
\caption{Parking occupancy prediction results under different distance thresholds.}
\label{tab:distance_result}
\resizebox{0.7\textwidth}{!}{
\begin{tabular}{c|c|ccc|ccc}
\hline
\multirow{2}{2cm}{Time}&\multirow{2}{1.0cm}{Metric} & \multicolumn{3}{c}{Weekday}&\multicolumn{3}{c}{Weekend}\\
\cline{3-8}
&&500 m & 1 km &Infinity &500 m & 1 km &Infinity\\
\hline
\multirow{3}{2cm}{Before COVID-19}&RMSE&0.07&0.41&3.18&0.06&0.16&2.63\\
&MAE &0.05&0.40&3.13&0.06&0.15&2.62\\
&MAPE &10.74 &29.18&682.50&12.62&22.37&422.57\\
\hline
\multirow{3}{2cm}{During COVID-19}&RMSE&0.09&0.11&4.50&0.28&0.43&1.18\\
&MAE &0.06&0.08&4.50&0.11&0.25&1.10\\
&MAPE &11.82 &19.73&677.60&12.22&39.30&196.64\\
\hline
\end{tabular}}
\end{table*}

To go deeper, we select two typical parking lots (i.e. Shipu restaurant and Shanjing shopping centre) to investigate into the performance of our framework and baseline GRU during COVID-19. In particular, this two typical parking lots are highly influenced by COVID-19 and have poor time regularities during this time period. The forecasting results are presented in Fig. \ref{fig:parking_covid_time}.
From the results, we can see that our framework has much better performance than GRU on predicting future parking occupancy during COVID-19. This further reinforce our findings that the incorporation of spatial dependence can help when there is less temporal regularity.
We therefore argue the importance of modelling spatial correlations for more accurate and robust parking occupancy prediction, especially during emergencies.

\subsection{Impact of Distance Threshold}
To evaluate the impact of distance threshold on forecasting accuracy, we compare the results of using different threshold settings, i.e. 500 m, 1 km, and infinity (full connection). The results are presented in Table \ref{tab:distance_result}.
As can be seen, the threshold of 500 m has the best performance for all the cases. The results imply that parking lots locating within 500 m have strong spatial correlations. This finding is consistent with previous studies in that passengers normally would prefer to walk up to 500 m \cite{gong2017geographical,yue2012exploratory}.

\subsection{Parking-related Activities Analysis}
\begin{figure}
\centering
\includegraphics[width=0.5\textwidth]{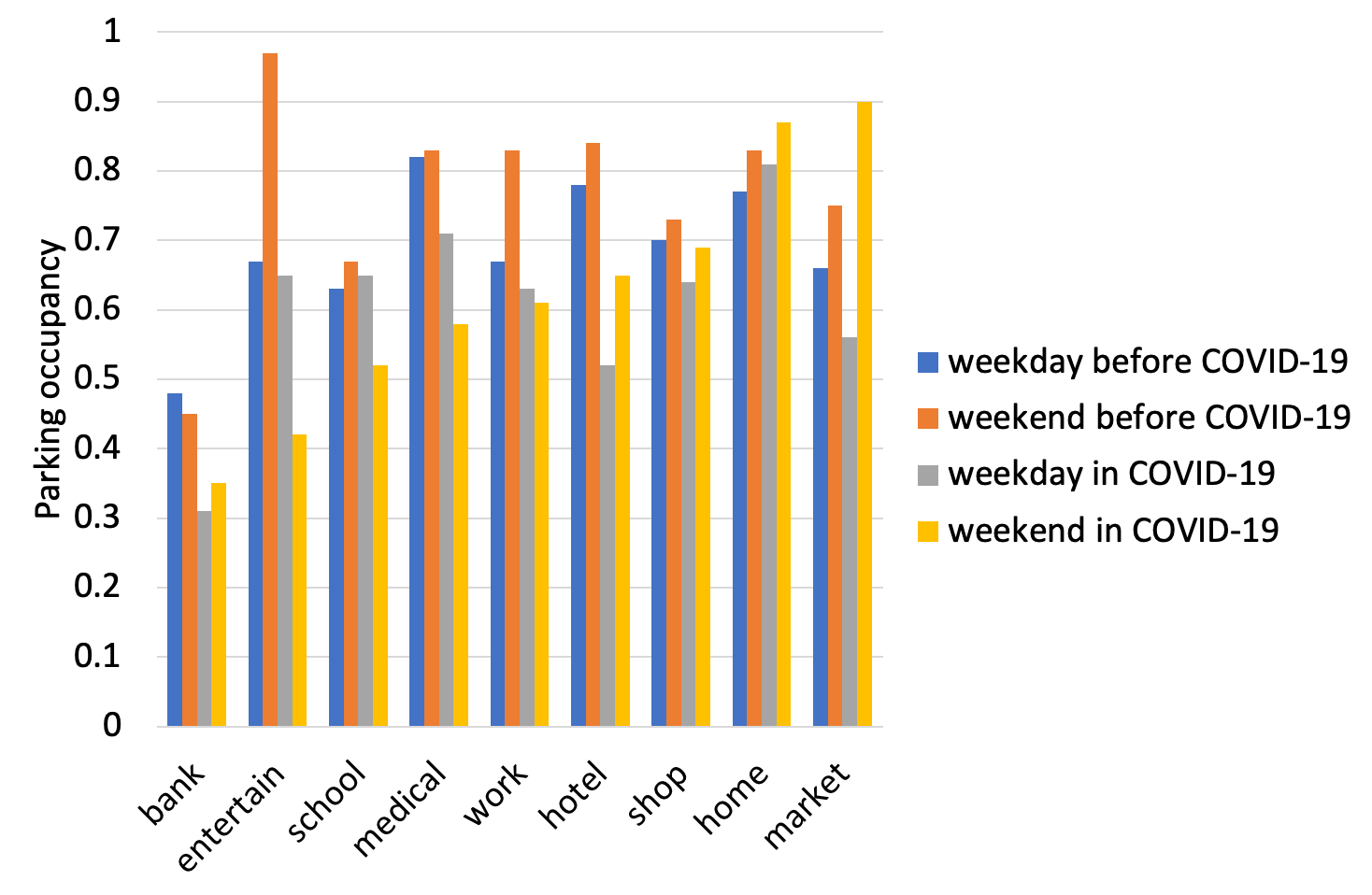}
\vspace{-0.5cm}
\caption{Parking occupancy of different activities in different times. }
\label{fig:parking_activity}
\end{figure}

To further investigate the reason why citizens' parking behaviours exhibit higher temporal irregularity, we extract passengers' social activity-related parking patterns between normal days and COVID-19 period.
Firstly, we annotate the related activity of each parking lot via Gaode Map (a Chinese online mapping service like Google Maps). Then, for each type of activity, we calculate the average occupancy of all related parking lots in the certain time period. The results are shown in Fig.~\ref{fig:parking_activity}. 

From the results, we can see that:
(i) In general, parking occupancy during COVID-19 is lower than normal days, especially for work, hotel, and shopping-related activities. This suggests that people reduce their time for unnecessary activities to avoid infection. One interpretation is, some citizens choose to work at home, and reduce the number of shopping time;
(ii) in weekdays, people would spent much more time to bank (48\% parking occupancy), medical institutions (82\%), taking a trip in different city (hotel-related activity, 78\%), or go to market in normal days than in COVID-19.
(iii) in weekends, citizens would spent less time especially for entertainment, school, and medical related activities during COVID-19. However, during COVID-19, market-related activities were much higher than normal (90\%). This is reasonable and suggests that, during COVID-19, people would spend as much time as they could staying at home to avoid infection, and they would prefer to buy sufficient daily necessities at weekends.
%
(iv) In addition, we find that parking lots which locate near the banks always have lower occupancy compared with other activities. This may be due to several reasons: first, banks reserve some parking space for the safety of cash truck; second, some banks would reserve parking space for their customers, and other cars are not welcomed.




\section{Conclusion}
\label{sec:conclusion}
In this paper, we have proposed a spatial-aware parking prediction framework leveraging both spatial and temporal correlations among parking lots. A case study in Ningbo, China has been conducted to evaluate the proposed method. Results demonstrate that the framework has higher predictive power than other baseline methods, especially for the case with less time regularities (such as during COVID-19). Our study therefore highlights the importance of spatial correlation, in addition to temporal dependence, in forecasting parking behaviours, particular in unexpected situations with irregular time patterns. Though proposed for parking prediction, our method can also be applied to make spatio-temporal forecasting of other location-related variables such as temperature and air quality from temporal observations of monitoring stations. 

In addition, we have also discovered spatio-temporal patterns of parking behaviour before and during COVID-19 in Ningbo, using parking data containing approximately one million parking records over half year time span. The results show that personal cars are used much less during COVID-19 than before, especially for unnecessary activities like entertainment and long-distance travelling.
Besides, most of the cars are parked in the residential areas during the pandemic, while passengers prefer to park their car in workplaces and residential areas in weekdays and city centres for entertainment in weekends before COVID-19.

In future, we would like to perform the following extensions: (i) link the parking data with more data sources to enhance the parking behaviour forecasting and analysis,
(ii) further improve the forecasting method by improving the integration of the connection graph and spatio-temporal forecasting model.



\section{Acknowledgments}
This work is supported by the Fundamental Research Funds for the Central Universities (Grant No.590121033), National Natural Science Foundation of China (Grant No.72071116), Natural Science Foundation of Zhejiang Province (Grant No. LR17G010001) and Ningbo Municipal Bureau of Science and Technology (Grant No. 2019B10026 \& 2017D10034).




\bibliographystyle{ACM-Reference-Format}




 \bibliography{main}


\begin{thebibliography}{20}


\ifx \showCODEN    \undefined \def \showCODEN     #1{\unskip}     \fi
\ifx \showDOI      \undefined \def \showDOI       #1{#1}\fi
\ifx \showISBNx    \undefined \def \showISBNx     #1{\unskip}     \fi
\ifx \showISBNxiii \undefined \def \showISBNxiii  #1{\unskip}     \fi
\ifx \showISSN     \undefined \def \showISSN      #1{\unskip}     \fi
\ifx \showLCCN     \undefined \def \showLCCN      #1{\unskip}     \fi
\ifx \shownote     \undefined \def \shownote      #1{#1}          \fi
\ifx \showarticletitle \undefined \def \showarticletitle #1{#1}   \fi
\ifx \showURL      \undefined \def \showURL       {\relax}        \fi
\providecommand\bibfield[2]{#2}
\providecommand\bibinfo[2]{#2}
\providecommand\natexlab[1]{#1}
\providecommand\showeprint[2][]{arXiv:#2}

\bibitem[\protect\citeauthoryear{Alajali, Wen, and Zhou}{Alajali
  et~al\mbox{.}}{2017}]%
        {alajali2017street}
\bibfield{author}{\bibinfo{person}{Walaa Alajali}, \bibinfo{person}{Sheng Wen},
  {and} \bibinfo{person}{Wanlei Zhou}.} \bibinfo{year}{2017}\natexlab{}.
\newblock \showarticletitle{On-street car parking prediction in smart city: a
  multi-source data analysis in sensor-cloud environment}. In
  \bibinfo{booktitle}{\emph{International Conference on Security, Privacy and
  Anonymity in Computation, Communication and Storage}}. Springer,
  \bibinfo{pages}{641--652}.
\newblock


\bibitem[\protect\citeauthoryear{Burns and Faurot}{Burns and Faurot}{1992}]%
        {burns1992econometric}
\bibfield{author}{\bibinfo{person}{Malcolm~R Burns} {and}
  \bibinfo{person}{David~J Faurot}.} \bibinfo{year}{1992}\natexlab{}.
\newblock \showarticletitle{An econometric forecasting model of revenues from
  urban parking facilities}.
\newblock \bibinfo{journal}{\emph{Journal of Economics and Business}}
  \bibinfo{volume}{44}, \bibinfo{number}{2} (\bibinfo{year}{1992}),
  \bibinfo{pages}{143--150}.
\newblock


\bibitem[\protect\citeauthoryear{Fiez, Ratliff, Dowling, and Zhang}{Fiez
  et~al\mbox{.}}{2018}]%
        {fiez2018data}
\bibfield{author}{\bibinfo{person}{Tanner Fiez}, \bibinfo{person}{Lillian~J
  Ratliff}, \bibinfo{person}{Chase Dowling}, {and} \bibinfo{person}{Baosen
  Zhang}.} \bibinfo{year}{2018}\natexlab{}.
\newblock \showarticletitle{Data driven spatio-temporal modeling of parking
  demand}. In \bibinfo{booktitle}{\emph{2018 Annual American Control Conference
  (ACC)}}. IEEE, \bibinfo{pages}{2757--2762}.
\newblock


\bibitem[\protect\citeauthoryear{Gong, Cartlidge, Bai, Yue, Li, and Qiu}{Gong
  et~al\mbox{.}}{2019}]%
        {gong2019extracting}
\bibfield{author}{\bibinfo{person}{Shuhui Gong}, \bibinfo{person}{John
  Cartlidge}, \bibinfo{person}{Ruibin Bai}, \bibinfo{person}{Yang Yue},
  \bibinfo{person}{Qingquan Li}, {and} \bibinfo{person}{Guoping Qiu}.}
  \bibinfo{year}{2019}\natexlab{}.
\newblock \showarticletitle{Extracting activity patterns from taxi trajectory
  data: a two-layer framework using spatio-temporal clustering, Bayesian
  probability and Monte Carlo simulation}.
\newblock \bibinfo{journal}{\emph{International Journal of Geographical
  Information Science}} (\bibinfo{year}{2019}), \bibinfo{pages}{1--25}.
\newblock


\bibitem[\protect\citeauthoryear{Gong, Cartlidge, Yue, Qiu, Li, and Xin}{Gong
  et~al\mbox{.}}{2017}]%
        {gong2017geographical}
\bibfield{author}{\bibinfo{person}{Shuhui Gong}, \bibinfo{person}{John
  Cartlidge}, \bibinfo{person}{Yang Yue}, \bibinfo{person}{Guoping Qiu},
  \bibinfo{person}{Qingquan Li}, {and} \bibinfo{person}{Jingyu Xin}.}
  \bibinfo{year}{2017}\natexlab{}.
\newblock \showarticletitle{Geographical huff model calibration using taxi
  trajectory data}. In \bibinfo{booktitle}{\emph{Proceedings of the 10th ACM
  SIGSPATIAL Workshop on Computational Transportation Science}}. ACM,
  \bibinfo{pages}{30--35}.
\newblock


\bibitem[\protect\citeauthoryear{Ji, Tang, Blythe, Guo, and Wang}{Ji
  et~al\mbox{.}}{2014}]%
        {ji2014short}
\bibfield{author}{\bibinfo{person}{Yanjie Ji}, \bibinfo{person}{Dounan Tang},
  \bibinfo{person}{Phil Blythe}, \bibinfo{person}{Weihong Guo}, {and}
  \bibinfo{person}{Wei Wang}.} \bibinfo{year}{2014}\natexlab{}.
\newblock \showarticletitle{Short-term forecasting of available parking space
  using wavelet neural network model}.
\newblock \bibinfo{journal}{\emph{IET Intelligent Transport Systems}}
  \bibinfo{volume}{9}, \bibinfo{number}{2} (\bibinfo{year}{2014}),
  \bibinfo{pages}{202--209}.
\newblock


\bibitem[\protect\citeauthoryear{Kipf and Welling}{Kipf and Welling}{2017}]%
        {kipf_semi-supervised_2017}
\bibfield{author}{\bibinfo{person}{Thomas~N. Kipf} {and} \bibinfo{person}{Max
  Welling}.} \bibinfo{year}{2017}\natexlab{}.
\newblock \showarticletitle{Semi-{Supervised} {Classification} with {Graph}
  {Convolutional} {Networks}}. In \bibinfo{booktitle}{\emph{5th {International}
  {Conference} on {Learning} {Representations}, {ICLR} 2017, {Toulon},
  {France}, {April} 24-26, 2017, {Conference} {Track} {Proceedings}}}.
\newblock


\bibitem[\protect\citeauthoryear{Klappenecker, Lee, and Welch}{Klappenecker
  et~al\mbox{.}}{2014}]%
        {klappenecker2014finding}
\bibfield{author}{\bibinfo{person}{Andreas Klappenecker},
  \bibinfo{person}{Hyunyoung Lee}, {and} \bibinfo{person}{Jennifer~L Welch}.}
  \bibinfo{year}{2014}\natexlab{}.
\newblock \showarticletitle{Finding available parking spaces made easy}.
\newblock \bibinfo{journal}{\emph{Ad Hoc Networks}}  \bibinfo{volume}{12}
  (\bibinfo{year}{2014}), \bibinfo{pages}{243--249}.
\newblock


\bibitem[\protect\citeauthoryear{Liu, Guan, Yan, and Yin}{Liu
  et~al\mbox{.}}{2010}]%
        {liu2010unoccupied}
\bibfield{author}{\bibinfo{person}{Shixu Liu}, \bibinfo{person}{Hongzhi Guan},
  \bibinfo{person}{Hai Yan}, {and} \bibinfo{person}{Huanhuan Yin}.}
  \bibinfo{year}{2010}\natexlab{}.
\newblock \showarticletitle{Unoccupied parking space prediction of chaotic time
  series}.
\newblock In \bibinfo{booktitle}{\emph{ICCTP 2010: Integrated Transportation
  Systems: Green, Intelligent, Reliable}}. \bibinfo{pages}{2122--2131}.
\newblock


\bibitem[\protect\citeauthoryear{Organization et~al\mbox{.}}{Organization
  et~al\mbox{.}}{2020}]%
        {world2020coronavirus}
\bibfield{author}{\bibinfo{person}{World~Health Organization} {et~al\mbox{.}}}
  \bibinfo{year}{2020}\natexlab{}.
\newblock \showarticletitle{Coronavirus disease 2019 (COVID-19): situation
  report, 54}.
\newblock  (\bibinfo{year}{2020}).
\newblock


\bibitem[\protect\citeauthoryear{Pullola, Atrey, and El~Saddik}{Pullola
  et~al\mbox{.}}{2007}]%
        {pullola2007towards}
\bibfield{author}{\bibinfo{person}{Sherisha Pullola},
  \bibinfo{person}{Pradeep~K Atrey}, {and} \bibinfo{person}{Abdulmotaleb
  El~Saddik}.} \bibinfo{year}{2007}\natexlab{}.
\newblock \showarticletitle{Towards an intelligent GPS-based vehicle navigation
  system for finding street parking lots}. In \bibinfo{booktitle}{\emph{2007
  IEEE International Conference on Signal Processing and Communications}}.
  IEEE, \bibinfo{pages}{1251--1254}.
\newblock


\bibitem[\protect\citeauthoryear{Tamrazian, Qian, and Rajagopal}{Tamrazian
  et~al\mbox{.}}{2015}]%
        {tamrazian2015my}
\bibfield{author}{\bibinfo{person}{Arbi Tamrazian}, \bibinfo{person}{Zhen
  Qian}, {and} \bibinfo{person}{Ram Rajagopal}.}
  \bibinfo{year}{2015}\natexlab{}.
\newblock \showarticletitle{Where is my parking spot? online and offline
  prediction of time-varying parking occupancy}.
\newblock \bibinfo{journal}{\emph{Transportation Research Record}}
  \bibinfo{volume}{2489}, \bibinfo{number}{1} (\bibinfo{year}{2015}),
  \bibinfo{pages}{77--85}.
\newblock


\bibitem[\protect\citeauthoryear{Tu, Cao, Yue, Zhou, Li, and Li}{Tu
  et~al\mbox{.}}{2018}]%
        {tu2018spatial}
\bibfield{author}{\bibinfo{person}{Wei Tu}, \bibinfo{person}{Rui Cao},
  \bibinfo{person}{Yang Yue}, \bibinfo{person}{Baoding Zhou},
  \bibinfo{person}{Qiuping Li}, {and} \bibinfo{person}{Qingquan Li}.}
  \bibinfo{year}{2018}\natexlab{}.
\newblock \showarticletitle{Spatial variations in urban public ridership
  derived from GPS trajectories and smart card data}.
\newblock \bibinfo{journal}{\emph{Journal of Transport Geography}}
  \bibinfo{volume}{69} (\bibinfo{year}{2018}), \bibinfo{pages}{45--57}.
\newblock


\bibitem[\protect\citeauthoryear{Vlahogianni, Kepaptsoglou, Tsetsos, and
  Karlaftis}{Vlahogianni et~al\mbox{.}}{2016}]%
        {vlahogianni2016real}
\bibfield{author}{\bibinfo{person}{Eleni~I Vlahogianni},
  \bibinfo{person}{Konstantinos Kepaptsoglou}, \bibinfo{person}{Vassileios
  Tsetsos}, {and} \bibinfo{person}{Matthew~G Karlaftis}.}
  \bibinfo{year}{2016}\natexlab{}.
\newblock \showarticletitle{A real-time parking prediction system for smart
  cities}.
\newblock \bibinfo{journal}{\emph{Journal of Intelligent Transportation
  Systems}} \bibinfo{volume}{20}, \bibinfo{number}{2} (\bibinfo{year}{2016}),
  \bibinfo{pages}{192--204}.
\newblock


\bibitem[\protect\citeauthoryear{Yang, Ma, Pi, and Qian}{Yang
  et~al\mbox{.}}{2019}]%
        {yang2019deep}
\bibfield{author}{\bibinfo{person}{Shuguan Yang}, \bibinfo{person}{Wei Ma},
  \bibinfo{person}{Xidong Pi}, {and} \bibinfo{person}{Sean Qian}.}
  \bibinfo{year}{2019}\natexlab{}.
\newblock \showarticletitle{A deep learning approach to real-time parking
  occupancy prediction in transportation networks incorporating multiple
  spatio-temporal data sources}.
\newblock \bibinfo{journal}{\emph{Transportation Research Part C: Emerging
  Technologies}}  \bibinfo{volume}{107} (\bibinfo{year}{2019}),
  \bibinfo{pages}{248--265}.
\newblock


\bibitem[\protect\citeauthoryear{Yue, Wang, Hu, Li, Li, and Yeh}{Yue
  et~al\mbox{.}}{2012}]%
        {yue2012exploratory}
\bibfield{author}{\bibinfo{person}{Yang Yue}, \bibinfo{person}{Han-dong Wang},
  \bibinfo{person}{Bo Hu}, \bibinfo{person}{Qing-quan Li},
  \bibinfo{person}{Yu-guang Li}, {and} \bibinfo{person}{Anthony~GO Yeh}.}
  \bibinfo{year}{2012}\natexlab{}.
\newblock \showarticletitle{Exploratory calibration of a spatial interaction
  model using taxi {GPS} trajectories}.
\newblock \bibinfo{journal}{\emph{Computers, Environment and Urban Systems}}
  \bibinfo{volume}{36}, \bibinfo{number}{2} (\bibinfo{year}{2012}),
  \bibinfo{pages}{140--153}.
\newblock


\bibitem[\protect\citeauthoryear{{Zhao}, {Song}, {Zhang}, {Liu}, {Wang}, {Lin},
  {Deng}, and {Li}}{{Zhao} et~al\mbox{.}}{2020}]%
        {zhao2019t}
\bibfield{author}{\bibinfo{person}{L. {Zhao}}, \bibinfo{person}{Y. {Song}},
  \bibinfo{person}{C. {Zhang}}, \bibinfo{person}{Y. {Liu}}, \bibinfo{person}{P.
  {Wang}}, \bibinfo{person}{T. {Lin}}, \bibinfo{person}{M. {Deng}}, {and}
  \bibinfo{person}{H. {Li}}.} \bibinfo{year}{2020}\natexlab{}.
\newblock \showarticletitle{T-GCN: A Temporal Graph Convolutional Network for
  Traffic Prediction}.
\newblock \bibinfo{journal}{\emph{IEEE Transactions on Intelligent
  Transportation Systems}} \bibinfo{volume}{21}, \bibinfo{number}{9}
  (\bibinfo{year}{2020}), \bibinfo{pages}{3848--3858}.
\newblock


\bibitem[\protect\citeauthoryear{Zheng, Rajasegarar, and Leckie}{Zheng
  et~al\mbox{.}}{2015}]%
        {zheng2015parking}
\bibfield{author}{\bibinfo{person}{Yanxu Zheng}, \bibinfo{person}{Sutharshan
  Rajasegarar}, {and} \bibinfo{person}{Christopher Leckie}.}
  \bibinfo{year}{2015}\natexlab{}.
\newblock \showarticletitle{Parking availability prediction for sensor-enabled
  car parks in smart cities}. In \bibinfo{booktitle}{\emph{2015 IEEE Tenth
  International Conference on Intelligent Sensors, Sensor Networks and
  Information Processing (ISSNIP)}}. IEEE, \bibinfo{pages}{1--6}.
\newblock


\bibitem[\protect\citeauthoryear{Zhou, Xu, Hu, Yue, Li, and Xia}{Zhou
  et~al\mbox{.}}{2020}]%
        {zhou2020effects}
\bibfield{author}{\bibinfo{person}{Ying Zhou}, \bibinfo{person}{Renzhe Xu},
  \bibinfo{person}{Dongsheng Hu}, \bibinfo{person}{Yang Yue},
  \bibinfo{person}{Qingquan Li}, {and} \bibinfo{person}{Jizhe Xia}.}
  \bibinfo{year}{2020}\natexlab{}.
\newblock \showarticletitle{Effects of human mobility restrictions on the
  spread of COVID-19 in Shenzhen, China: a modelling study using mobile phone
  data}.
\newblock \bibinfo{journal}{\emph{The Lancet Digital Health}}
  \bibinfo{volume}{2}, \bibinfo{number}{8} (\bibinfo{year}{2020}),
  \bibinfo{pages}{e417--e424}.
\newblock


\bibitem[\protect\citeauthoryear{Ziat, Leroy, Baskiotis, and Denoyer}{Ziat
  et~al\mbox{.}}{2016}]%
        {ziat2016joint}
\bibfield{author}{\bibinfo{person}{Ali Ziat}, \bibinfo{person}{Bertrand Leroy},
  \bibinfo{person}{Nicolas Baskiotis}, {and} \bibinfo{person}{Ludovic
  Denoyer}.} \bibinfo{year}{2016}\natexlab{}.
\newblock \showarticletitle{Joint prediction of road-traffic and parking
  occupancy over a city with representation learning}. In
  \bibinfo{booktitle}{\emph{2016 IEEE 19th International Conference on
  Intelligent Transportation Systems (ITSC)}}. IEEE, \bibinfo{pages}{725--730}.
\newblock


\end{thebibliography}
\end{document}